\documentclass[11pt,a4paper]{article}
\usepackage{jheppub}
\usepackage{amsmath,amssymb,graphics,epsfig,color,mathrsfs,cases,colortbl,framed}

\title{{ $U(1)$ gauge vector field on a codimension-2 brane}}
\author{Chun-E Fu$^{\dag}$, Yuan Zhong$^{\dag}$, and Yu-Xiao Liu$^{\ddag}$}
\affiliation[]{
$^{\dag}$School of Science, Xi'an Jiaotong University, Xi'an 710049, China\\
$^{\ddag}$Institute of Theoretical Physics $\&$ Research Center of Gravitation, Lanzhou University, Lanzhou 730000, China\\
}
\emailAdd{fuche13@mail.xjtu.edu.cn}
\emailAdd{zhongy@mail.xjtu.edu.cn}
\emailAdd{liuyx@lzu.edu.cn}

\abstract{ In this paper, we obtain a gauge invariant effective action for a bulk massless  $U(1)$ gauge vector field on a brane with codimension two by using a general Kaluza-Klein (KK) decomposition for the field. It suggests that there exist two types of scalar KK modes to keep the gauge invariance of the action for the massive vector KK modes. Both the vector and scalar KK modes can be massive. The masses of the vector KK modes $m^{(n)}$ contain two parts, $m_{1}^{(n)}$ and $m_{2}^{(n)}$, due to the existence of the two extra dimensions. The masses of the two types of scalar KK modes $m_{\phi}^{(n)}$ and $m_{\varphi}^{(n)}$ are related to the vector ones, i.e., $m_{\phi}^{(n)}=m_{1}^{(n)}$ and  $m_{\varphi}^{(n)}=m_{2}^{(n)}$. Moreover, we derive two Schr\"{o}dinger-like equations for the vector KK modes, for which the effective potentials are just the functions of the warp factor.
}

\begin{document}
\maketitle

\section{Introduction}

If there are extra dimensions, the physical world will be surely more interesting. For example, in the {Kaluza-Klein (KK) theory with one extra spatial dimension,} the {four-dimensional} (4D) gravity and electromagnetism can be unified. In the Arkani-Hamed-Dimopoulos-Dvali (ADD) theory \cite{ArkaniHamed1998rs}, the hierarchy problem is reconsidered {to be related with large extra dimensions.} Especially, in the Randall-Sundrum (RS) brane world theory \cite{Randall1999a,Randall1999b}, the warped space-time outside our {four-dimensional} world leads to some new and dramatic phenomenologies. That's why the extra dimension and brane world theories are paid for more and more attention \cite{coscon2009,ThickBrane2002,Cline:2003ak,PhysRevD.66.024024,Bazeia2015owa,Archer2011,Gani:2016fqq,
Yang:2017evd,Liu2017gcn,Zhong2017uhn,Parameswaran2009bt}.

{One of the most interesting things is that there will be KK modes for various bulk fields with extra dimensions.} In the brane world theory, the zero modes can be {regarded as} particles living in four-dimensional {space-time, while} the massive KK modes might reveal the mysteries {of extra dimensions. Thus,} there has been many papers focus on these KK modes \cite{Chang1999nh,RandjbarDaemi:2000cr,Oda2001,Mukhopadhyaya:2001fc,Ichinose:2002kg,
Gogberashvili:2007gg,Guo:2008ia,Liu2008WeylPT,Liu:2009mga,LocalizationWithoutScalar2011JHEP,
Fu2012sa,Fu2015cfa,Guo2014nja,Vaquera-Araujo2014tia,Arun:2016ela,
Li2017dkw,Mendes2017hmv,Alencar2018cbk,Zhou2017bbj}.

In this paper, we will discuss the KK modes of $U(1)$ gauge vector field in a brane model with codimension two. It is known that the localization of the vector field is difficult in five-dimensional brane world model, and so there are many literatures in order to find localization mechanism \cite{Zhao2014gka,Zhao2014iqa,Alencar2014moa,Zhao2017epp,Kehagias:2000au,Chumbes2011zt,
Ghoroku:2001zu,Freitas2018iil}. In this paper, would like to think about another question for the KK modes of a vector field, i.e., how to keep the invariance of the brane effective action. It may be argued that the invariance of the action can be guaranteed by the Higgs mechanism. But here we begin with a massless vector field in the bulk space-time, and only consider the influence of the extra dimensions on the KK modes.

For the zero modes, the action is straightforward gauge invariant not only in the bulk but also on the brane. For the massive vector KK modes, it is not easy to obtain a gauge invariant effective action on the brane. Fortunately, in our recently work, we have investigated a new localization mechanism of an arbitrary $q-$form field on $p-$brane with codimension one \cite{Fu2015cfa}. We discovered that with a general KK decomposition for the bulk $q-$form field there is a gauge invariant effective action for the KK modes. For example, by the new localization mechanism for a $1-$form bulk field (vector field), we can get a gauge invariant effective action for the massive vector KK modes, which are coupled with some massless scalar KK modes. This is similar to the Higgs mechanism which enables a massless vector to obtain mass. The difference is that in this brane world model with codimension one, the vector KK modes obtain masses due to the existence of extra dimension.

Therefore, it is reasonable to think that on a brane with codimension two the vector KK modes would get two parts of masses from the two extra dimensions. In this work, we would like to verify this idea. What interesting is that we also obtain a gauge invariance effective action on the brane. This effective action indicates that the vector KK modes really obtain their masses from the two extra dimensions. In addition, there are two types of massive scalar KK modes.

Our discussion is based on the following line-element:
\begin{equation}
  ds^2=\text{e}^{2A(y,z)}\big(\hat{g}_{\mu\nu} dx^\mu\ dx^\nu+dy^2 +dz^2\big),\label{line-element1}
\end{equation}
 where the warp factor $A(y,z)$ is a function of the two extra dimensional coordinates $y$ and $z$, and $\hat{g}_{\mu\nu}$ is the induced metric on the brane. The bulk fields can propagate along the two directions outside the brane, and play the role of KK modes on the brane. In the following section we will start our work with a general KK decomposition for the bulk vector field. Then we will try to find two groups of equations of motion (EOM) for the KK modes of the field, and will discover some interesting results by comparing these EOMs.

\section{KK decomposition and the effective action}\label{localizationmechanism}

{ In this paper, we consider the model of brane world with codimension two. The line-element is given by \eqref{line-element1}. The} action for a bulk massless { $U(1)$} gauge vector field is
\begin{eqnarray}\label{bulkaction}
  S&=&-\frac{1}{4}\int d^6 x\sqrt{-g}\;Y^{M_1M_2}Y_{M_1M_2}\nonumber\\
  &=&-\frac{1}{4}\int d^6 x\sqrt{-g}\bigg(Y^{\mu_1\mu_2}Y_{\mu_1\mu_2}+2Y^{\mu_1z}Y_{\mu_1z}
  +2Y^{\mu_1y}Y_{\mu_1y}+2Y^{yz}Y_{yz}\bigg),
\end{eqnarray}
where $Y_{M_1M_2}=\frac{1}{2}\big(\partial_{M_1}X_{M_2}-\partial_{M_2}X_{M_1}\big)$
is the field strength of the vector field $X_{M}$. The equations of motion for the bulk field are given by
\begin{eqnarray}
  \partial_{\mu_1}(\sqrt{-g}Y^{\mu_1\mu_2})+ \partial_{z}(\sqrt{-g}Y^{z\mu_2})+ \partial_{y}(\sqrt{-g}Y^{y\mu_2})&=&0,\label{equ1}\\
 \partial_{\mu_1}(\sqrt{-g}Y^{\mu_1z})+ \partial_{y}(\sqrt{-g}Y^{yz})&=&0,\label{equ2}\\
 \partial_{\mu_1}(\sqrt{-g}Y^{\mu_1y})+ \partial_{z}(\sqrt{-g}Y^{zy})&=&0.\label{equ3}
\end{eqnarray}

This action \eqref{bulkaction} is surely gauge invariant. { With a KK decomposition of the vector field and a dimension reduction, we will obtain the effective actions for the vector KK modes. It is easy to show that the effective action of the vector zero mode is also gauge invariant, while the ones for the massive KK modes are not without some mechanisms \cite{Liu2007WeylVo,Liu2009uca,LocalizationFuPRD2011,Costa:2013eua}.} 

We {have found} one way to get the gauge invariant actions for the massive vector KK modes {on} a brane with codimension one, where our aim was in fact to solve the Hodge duality on the brane for the $q-$form field \cite{Fu2015cfa}. We know that, { with a given KK decomposition of a bulk field, the effective actions on the brane} for the KK modes of the field can be finally derived. {Different KK decompositions will lead to different effective actions for the KK modes. Usually, one would like to choose a gauge to simplify the calculation before doing the KK decomposition for the vector field \cite{Duff2000se}.} We think that the choice of the gauge for the vector field in fact influences the KK decomposition. Therefore, in order to obtain a gauge invariant effective brane action, one should start with a general KK decomposition without using a gauge fixing.

Thus, we begin with a general KK decomposition for the bulk massless {vector field:}
\begin{eqnarray}
  X_{\mu_1}(x_{\mu},y,z)&=&\sum_{n}\hat{X}_{\mu_1}^{(n)}(x^{\mu})
  \;W_1^{(n)}(y,z)\;\text{e}^{a_1\;A(y,z)}, \label{kk1} \\
  X_z(x_{\mu},y,z)&=&\sum_{n}\phi^{(n)}(x^{\mu})
  \;W_2^{(n)}(y,z)\;\text{e}^{a_2\;A(y,z)}, \label{kk2}\\
  X_y(x_{\mu},y,z)&=&\sum_{n}\varphi^{(n)}(x^{\mu})
  \;W_3^{(n)}(y,z)\;\text{e}^{a_3\;A(y,z)}, \label{kk3}
\end{eqnarray}
where { $W_1^{(n)}(y,z)$, $W_2^{(n)}(y,z)$, and $W_3^{(n)}(y,z)$} are only the functions of extra dimensions $y, z$, and $a_1$, $a_2$, and $a_3$ can be taken as any arbitrary constants. Note that we always use $(n)$ and $(n')$ to label KK modes in this paper.

With the above KK decompositions, we can {derive} the effective actions for the KK modes { $\hat{X}_{\mu_1}^{(n)}(x^{\mu})$, $\phi^{(n)}(x^{\mu})$, and $\varphi^{(n)}(x^{\mu})$}:
\begin{eqnarray}
  S&=&-\frac{1}{4}\int {d^6 x} \sqrt{-g}\;Y^{M_1M_2}Y_{M_1M_2}\nonumber\\
  &=&-\frac{1}{4}\int {d^6 x} \sqrt{-g}\bigg(Y^{\mu_1\mu_2}Y_{\mu_1\mu_2}+2Y^{\mu_1z}Y_{\mu_1z}
  +2Y^{\mu_1y}Y_{\mu_1y}+2Y^{yz}Y_{yz}\bigg),
  \nonumber\\
  \nonumber\\
   &=&
     -\frac{1}{4}\sum_n\sum_{n'}\int {d^{4}x} \sqrt{-\hat{g}}
     \;\bigg[
            I_{1}^{(nn')}\;\hat{Y}^{(n)}_{\mu_1\mu_2}\;\hat{Y}^{\mu_1\mu_2(n')}
            +\big(I_{2}^{(nn')}+I_{4}^{(nn')}\big)\;\hat{X}_{\mu_1}^{(n)}\hat{X}^{\mu_1(n')}\nonumber\\
          &&+I_{3}^{(nn')}\;\partial_{\mu_1}\phi^{(n)}\;\partial^{\mu_1}\phi^{(n')}
            -I_{6}^{(nn')}\;\bigg(\partial_{\mu_1}\phi^{(n)}\;\hat{X}^{\mu_1(n')}
                                +\hat{X}_{\mu_1}^{(n)}\;\partial^{\mu_1}\phi^{(n')}\bigg)\nonumber\\
          &&+I_{5}^{(nn')}\;\partial_{\mu_1}\varphi^{(n)}\;\partial^{\mu_1}\varphi^{(n')}
            -I_{8}^{(nn')}\;\bigg(\partial_{\mu_1}\varphi^{(n)}\;\hat{X}^{\mu_1(n')}
                                +\hat{X}_{\mu_1}^{(n)}\;\partial^{\mu_1}\varphi^{(n')}\bigg)\nonumber\\
          &&+I_{7}^{(nn')}\;\phi^{(n)}\phi^{(n')}
            +I_{9}^{(nn')}\;\varphi^{(n)}\varphi^{(n')}
            -\;I_{10}^{(nn')}\;
            \bigg(\phi^{(n)}\varphi^{(n')}+\varphi^{(n)}\phi^{(n')}\bigg)
                                 \bigg],\label{effectiveAction}
\end{eqnarray}
where we have assumed that $W_1^{(n)}(y,z)$, $W_2^{(n)}(y,z)$, and $W_3^{(n)}(y,z)$ satisfy the following orthonormality conditions:
\begin{subequations}\label{OrthonormalityCondition}
\begin{eqnarray}
  I_{1}^{(nn')}&\equiv&\int dy\;dz\; W_1^{(n)}W_1^{(n')}
    =\delta_{nn'},\label{condition1}\\
 I_{3}^{(nn')}&\equiv&\frac{1}{2}\;\int dy\;dz\; W_2^{(n)}W_2^{(n')}
    =2\delta_{nn'},\label{condition3}\\
 I_{5}^{(nn')}&\equiv&\frac{1}{2}\;\int dy\;dz\; W_3^{(n)}W_3^{(n')}
    =2\delta_{nn'},\label{condition5}
\end{eqnarray}
\end{subequations}
and the other constants are given by
\begin{subequations}\label{constants}
\begin{eqnarray}
I_{2}^{(nn')}&\equiv&\frac{1}{2}\;\int dy\;dz\;
     \partial_y(W_1^{(n)}\text{e}^{-A})\;\partial_y(W_1^{(n')}\text{e}^{-A})
     \;\text{e}^{2A}<\infty , \label{condition2}\\
I_{4}^{(nn')} &\equiv&\frac{1}{2}\; \int dy\;dz\;
     \partial_z(W_1^{(n)}\text{e}^{-A})\;\partial_z(W_1^{(n')}\text{e}^{-A})
     \;\text{e}^{2A}<\infty  \label{condition4},\\
I_{6}^{(nn')}&\equiv& \frac{1}{2}\;\int dy\;dz\;
     W_2^{(n)}\;\partial_z(W_1^{(n')}\text{e}^{-A})\;\text{e}^{A}<\infty,  \label{condition6}\\
I_{8}^{(nn')} &\equiv&\frac{1}{2}\;\int dy\;dz\;
     W_3^{(n)}\;\partial_y(W_1^{(n')}\text{e}^{-A})\;\text{e}^{A}<\infty,  \label{condition8}\\
I_{7}^{(nn')} &\equiv& \frac{1}{2}\;\int dy\;dz\;
     \partial_y(W_2^{(n)}\text{e}^{-A})\partial_y(W_2^{(n')}\text{e}^{-A})
     \;\text{e}^{2A}<\infty,  \label{condition7}\\
I_{9}^{(nn')}&\equiv&\frac{1}{2}\;\int dy\;dz\;
     \partial_z(W_3^{(n)}\text{e}^{-A})\partial_z(W_3^{(n')}\text{e}^{-A})
     \;\text{e}^{2A}<\infty,  \label{condition9}\\
I_{10}^{(nn')} &\equiv&\frac{1}{2}\;\int dy\;dz\;
     \partial_y(W_2^{(n)}\text{e}^{-A})\;
     \partial_z(W_3^{(n)}\text{e}^{-A})
     \;\text{e}^{2A}<\infty.  \label{condition10}
\end{eqnarray}
\end{subequations}
We have taken $a_1=a_2=a_3=-1$, and defined $\hat{Y}^{(n)}_{\mu_1\mu_2}\equiv\frac{1}{2}\big(\partial_{\mu_1}\hat{X}_{\mu_2}^{(n)}
-\partial_{\mu_2}\hat{X}_{\mu_1}^{(n)}\big)$.

From the effective action \eqref{effectiveAction}, we can analyze the following useful information:
\begin{itemize}
  \item {There are three types of KK modes, the vector $\hat{X}_\mu^{(n)}$, and two types of scalar ones $\phi^{(n)}$, $\varphi^{(n)}$.}{   The constant symmetric matrixes ${2}(I_{2}^{(nn)}+I_{4}^{(nn)})$, $\frac{1}{2}I_{7}^{(nn)}$ and $\frac{1}{2} I_{9}^{(nn)}$ are the mass matrixes of the vector $\hat{X}_{\mu_1}^{(n)}(x^{\mu})$, and the scalar modes $\phi^{(n)}(x^{\mu})$ and $\varphi^{(n)}(x^{\mu})$, respectively;}
  \item The asymmetric matrixes $I_{6}^{(nn')}$ and $I_{8}^{(nn')}$ describe the coupling between the vector and scalar KK modes, and $I_{10}^{(nn')}$ between two {scalar modes};
  \item { In the effective action \eqref{effectiveAction}, except for the term $\hat{Y}^{(n)}_{\mu_1\mu_2}\;\hat{Y}^{\mu_1\mu_2(n')}$, if the other terms can be rewritten as the form
  \begin{equation}
    (\partial_{\mu}\phi^{(n)}-C\;\hat{X}_{\mu}^{(n)})^2, ~~
    (\partial_{\mu}\varphi^{(n)}-C\;\hat{X}_{\mu}^{(n)})^2,~~
    (\phi^{(n)}-\varphi^{(n)})^2,
  \end{equation}
  the action is gauge invariant under the transformation
  \begin{equation}
    \hat{X}_{\mu}^{(n)}\rightarrow \hat{X}_{\mu}^{(n)}+\partial_{\mu}\rho^{(n)}, ~~
    \phi^{(n)}\rightarrow \phi^{(n)}+C\rho^{(n)}, ~~
    \varphi^{(n)}\rightarrow \varphi^{(n)}+C\rho^{(n)},
  \end{equation}
  where $C$ is a constant and $\rho^{(n)}$ a scalar field.}
\end{itemize}

We can guess that if the effective action \eqref{effectiveAction} is really invariant, there must be some relationship between the mass parameters and the coupling constants. Through deriving the equations the KK modes satisfied, we will find these relationships.

\section{KK modes of the $U(1)$ gauge vector field  }

To derive the equations of the KK modes, we will compare two groups of EOMs for the KK modes. One group of EOMs is obtained from the effective action \eqref{effectiveAction}, and another is from substituting the KK decomposition into Eqs. \eqref{equ1}-\eqref{equ3}.

{ To simplify the calculation, we now introduce four mass parameters for the vector and scalar KK modes:
  \begin{eqnarray}
  I_{2}^{(nn)}&=&\frac{1}{2}\; {m_{1}^{(n)}}^2,\\
  I_{4}^{(nn)}&=&\frac{1}{2}\; {m_{2}^{(n)}}^{2},\\
  I_{7}^{(nn)}&=&2\; {m_\phi^{(n)}}^{2},\\
  I_{9}^{(nn)}&=&2\; {m_\varphi^{(n)}}^{2},
\end{eqnarray}}
where $ m_{v}^{(n)} \equiv \sqrt{m_{1}^{(n)2} + m_{2}^{(n)2}}$ are the masses of the vector KK modes, and $m_{\phi}^{(n)}$ and $m_{\varphi}^{(n)}$ are the masses of the two scalar modes $\phi^{(n)}$ and $\varphi^{(n)}$, respectively.

\subsection{{The equations of KK modes}}

Firstly, from the effective action \eqref{effectiveAction}, we obtain three equations:
\begin{eqnarray}
  &&\!\!\!\!\!\!\frac{1}{\sqrt{-\hat{g}}}\partial_{\mu_1}\left(I_{1}^{(nn')}\sqrt{-\hat{g}}\hat{Y}^{\mu_1\mu_2(n')}\right)
  -\big(I_{2}^{(nn')}-I_{4}^{(nn')}\big)\hat{X}^{\mu_2(n')}
  +I_{6}^{(nn')}\partial^{\mu_2}\phi^{(n')}
  +I_{8}^{(nn')}\partial^{\mu_2}\varphi^{(n')}=0, \nonumber\\\label{effequ1}\\
 && \frac{1}{\sqrt{-\hat{g}}}\partial_{\mu_1}
  \left(I_{3}^{(nn')}\sqrt{-\hat{g}}\partial^{\mu_1}\phi^{(n')}
  -I_{6}^{(nn')}\sqrt{-\hat{g}}\hat{X}^{\mu_1(n')}\right)
  -I_{7}^{(nn')}\phi^{(n')}+I_{10}^{(nn')}\varphi^{(n')}=0,\label{effequ2}\\
  &&\frac{1}{\sqrt{-\hat{g}}}\partial_{\mu_1}\left(I_{5}^{(nn')}\sqrt{-\hat{g}}\partial^{\mu_1}\varphi^{(n')}
  -I_{8}^{(nn')}\sqrt{-\hat{g}}\hat{X}^{\mu_1(n')}\right)
  -I_{9}^{(nn')}\varphi^{(n')}+I_{10}^{(nn')}\phi^{(n')}=0\label{effequ3}.
\end{eqnarray}

Secondly, by inserting the KK decomposition \eqref{kk1}-\eqref{kk3} into Eqs. \eqref{equ1}-\eqref{equ3}, we have
\begin{eqnarray}\label{effequ11}
  &&\!\!\!\!\!\!\!\!\frac{1}{\sqrt{-\hat{g}}}\;\partial_{\mu_1}
  \left(\sqrt{-\hat{g}}\;\hat{Y}^{\mu_1\mu_2(n)}\right)\;
  +\lambda_1\;\hat{X}^{\mu_2(n)}
  +\lambda_2\;\hat{X}^{\mu_2(n)}
  -\lambda_3\;\partial^{\mu_2}\varphi^{(n)}
  -\lambda_4\;\partial^{\mu_2}\phi^{(n)}=0 ,\\
  \label{effequ22}
 &&\frac{1}{\sqrt{-\hat{g}}}\partial_{\mu_1}
  \bigg(\sqrt{-\hat{g}}\;\partial^{\mu_1}\phi^{(n)}
  -\lambda_5\;\sqrt{-\hat{g}}\;\hat{X}^{\mu_1(n)}\bigg)
  +\lambda_6\;\phi^{(n)}
  -\lambda_7\;\varphi^{(n)}=0,\\
  \label{effequ33}
 &&\frac{1}{\sqrt{-\hat{g}}}\partial_{\mu_1}
  \bigg(\sqrt{-\hat{g}}\;\partial^{\mu_1}\varphi^{(n)}
  -\;\lambda_8\sqrt{-\hat{g}}\;\hat{X}^{\mu_1(n)}\bigg)
  -\lambda_9\;\phi^{(n)}
  +\lambda_{10}\;\varphi^{(n)}=0,
\end{eqnarray}
where we define ten coefficients:
\begin{eqnarray}
\lambda_1 &\equiv& \frac{\partial_y
  \left(\partial_y\big(W^{(n)}_1\;\text{e}^{-A}\big)
  \text{e}^{2A}\right)\;\text{e}^{-A}}
  {2W_1^{(n)}}, \quad
\lambda_3 \equiv \frac{\partial_y\big(W^{(n)}_3\text{e}^{A}\big)
  \;\text{e}^{-A}}{2W_1^{(n)}},\nonumber\\
\lambda_2&\equiv&\frac{\partial_z\bigg(\partial_z\big(W^{(n)}_1\;\text{e}^{-A}\big)
  \text{e}^{2A}\bigg)
  \text{e}^{-A}}{2W_1^{(n)}}, \quad ~
\lambda_4\equiv\frac{\partial_z\big(W^{(n)}_2\text{e}^{A}\big)
  \;\text{e}^{-A}} {2W_1^{(n)}},\nonumber\\
\lambda_6 &\equiv& \frac{\partial_y
  \left[
  \partial_y\big(W_2^{(n)}\;\text{e}^{-A}\big)
  \text{e}^{2A}\right]\text{e}^{-A}}{W_2^{(n)}}, \quad ~
\lambda_5\equiv\frac{
  \partial_z\big(W_1^{(n)}\;\text{e}^{-A}\big)
  \text{e}^{A}}{W_2^{(n)}},\nonumber
\end{eqnarray}
\begin{eqnarray}
\lambda_7&\equiv&\frac{\partial_y
  \left[
  \partial_z\big(W_3^{(n)}\;\text{e}^{-A}\big)
  \text{e}^{2A}\right]
  \text{e}^{-A}}{W_2^{(n)}},\quad
  \lambda_8\equiv\frac{
  \partial_y\big(W_1^{(n)}\;\text{e}^{-A}\big)
  \text{e}^{A}}{W_3^{(n)}},\nonumber\\
\lambda_9&\equiv&\frac{\partial_z
  \left[
  \partial_y\big(W_2^{(n)}\;\text{e}^{-A}\big)
  \text{e}^{2A}\right]\text{e}^{-A}}{W_3^{(n)}},\quad
\lambda_{10}\equiv\frac{\partial_z
  \left[
  \partial_z\big(W_3^{(n)}\;\text{e}^{-A}\big)
  \text{e}^{2A}\right]
  \text{e}^{-A}}{W_3^{(n)}}.\nonumber
\end{eqnarray}

Then we compare Eqs. \eqref{effequ1}-\eqref{effequ3} with \eqref{effequ11}- \eqref{effequ33}.
\begin{itemize}
  \item
We first focus on the terms containing $I_{2}^{(nn')}, I_{4}^{(nn')},I_{7}^{(nn')}$ and $I_{9}^{(nn')}$. Through the comparision, it is found
\begin{eqnarray}
I_{2}^{(nn')}&=&-\delta _{nn'}\lambda_1,\label{coup1}\\
I_{4}^{(nn')}&=&-\delta _{nn'}\lambda _2,\label{coup2}\\
I_{7}^{(nn')}&=&-2\delta _{nn'}\lambda _6,\label{coup3}\\
I_{9}^{(nn')}&=&-2\delta _{nn'}\lambda _{10},\label{coup4}
\end{eqnarray}
which leads to four Schr\"{o}dinger-like equations:
\begin{eqnarray}
  \big[-\partial_y^2+P_1\big]W^{(n)}_1(y,z) &=&m_{1}^{(n)2}\;W^{(n)}_1(y,z),\label{sch1}\\
  \big[-\partial_z^2+P_2\big]W^{(n)}_1(y,z) &=&m_{2}^{(n)2}\;W^{(n)}_1(y,z),\label{sch2}\\
  \big[-\partial_y^2+P_1\big]W^{(n)}_2(y,z) &=&m_{\phi}^{(n)2}\;W^{(n)}_2(y,z),\label{sch3}\\
  \big[-\partial_z^2+P_2\big]W^{(n)}_3(y,z) &=&m_{\varphi}^{(n)2}\;W^{(n)}_3(y,z),\label{sch4}
\end{eqnarray}
with $P_1(y)$ and $P_2(z)$ the effective potentials
\begin{eqnarray}
P_1(y)&=&\partial_y^2A(y,z) +\partial_yA(y,z)\;\partial_yA(y,z),\label{veff1}\\
P_2(z)&=&\partial_z^2A(y,z) +\partial_zA(y,z)\;\partial_zA(y,z).\label{veff2}
\end{eqnarray}
Given the solution of the background, the masses and wave functions of the vector KK modes can be solved through Eqs. \eqref{sch1} and \eqref{sch2}. But for the two types of scalar KK modes, the equations \eqref{sch3} and \eqref{sch4} only tell us the behaviors of their wave functions along one of the extra dimensions. Continuing to compare the two groups of EOMs, we will find more equations.

  \item
Considering the terms about $I_{6}^{(nn')}$, $I_{8}^{(nn')}$ and $I_{10}^{(nn')}$, it is also easy to get some equations from the comparison. But for convenience we introduce three constants $C_{1}^{(n)}, C_{2}^{(n)}, C_{3}^{(n)}$ to simplify the calculation, and let
\begin{eqnarray}
I_{6}^{(nn')}&=&C_{1}^{(n)}\;\delta_{n'n},\\
I_{8}^{(nn')}&=&C_{2}^{(n)}\;\delta_{n'n},\\
I_{10}^{(nn')}&=&C_{3}^{(n)}\;\delta_{n'n}.
\end{eqnarray}
So we can obtain
\begin{eqnarray}
 -2\;C_{1}^{(n)}\;{W_1^{(n)}}&=&
 \partial_z\big(W^{(n)}_2\text{e}^{A}\big)\;\text{e}^{-A},\label{coupl1}\\
 \frac{1}{2}\;C_{1}^{(n)}\;{W_2^{(n)}}&=&
 \partial_z\big(W_1^{(n)}\;\text{e}^{-A}\big)\text{e}^{A},\label{coupl2}\\
 -2\;C_{2}^{(n)}\;{W_1^{(n)}}&=&
  \partial_y\big(W^{(n)}_3\text{e}^{A}\big)
  \;\text{e}^{-A},\label{coupl3}\\
 \frac{1}{2}\;C_{2}^{(n)}\;{W_3^{(n)}}&=&
  \partial_y\big(W_1^{(n)}\;\text{e}^{-A}\big)
  \text{e}^{A},\label{coupl4}\\
 -\frac{1}{2}\;C_{3}^{(n)}\;{W_2^{(n)}}&=&\partial_y
  \left[
  \partial_z\big(W_3^{(n)}\;\text{e}^{-A}\big)
  \text{e}^{2A}\right]
  \text{e}^{-A}\label{coupl5},\\
 -\frac{1}{2}\;C_{3}^{(n)}\;{W_3^{(n)}}&=&
  \partial_z
  \left[
  \partial_y\big(W_2^{(n)}\;\text{e}^{-A}\big)
  \text{e}^{2A}\right]\text{e}^{-A}.
  \label{coupl6}
\end{eqnarray}
These equations show the relationships between the wave functions $W_1^{(n)}(y,z), W_2^{(n)}(y,z)$ and $W_3^{(n)}(y,z)$. We can further simplify them.
  \begin{itemize}
  \item

With the Schr\"{o}dinger-like equation \eqref{sch2} and Eq. \eqref{coupl2}, we have $ C_{1}^{(n)}=m_2^{(n)2}$. Then with \eqref{coupl1} we can find another Schr\"{o}dinger-like equation for $W_2(y,z)^{(n)}$. It is similar to $C_{2}^{(n)}$ and $W_3^{(n)}(y,z)$. We list the result:
\begin{eqnarray}
  C_{1}^{(n)2}&=&m_2^{(n)2},\label{rela1}\\
  C_{2}^{(n)2}&=&m_1^{(n)2},\label{rela2}
\end{eqnarray}
and
\begin{eqnarray}
\big[-\partial_z^2+P_3\big]W^{(n)}_2(y,z)&=&m_2^{(n)2}\;W^{(n)}_2(y,z),\label{sch5}\\
\big[-\partial_y^2+P_4\big]W^{(n)}_3(y,z)&=&m_1^{(n)2}\;W^{(n)}_3(y,z),\label{sch6}
\end{eqnarray}
where $P_3$ and $P_4$ are the effective potentials:
\begin{eqnarray}
P_3&=&\partial_zA\;\partial_zA-\partial_z^2A,\label{veff5}\\
P_4&=&\partial_yA\;\partial_yA-\partial_y^2A.\label{veff6}
\end{eqnarray}

Now there are two more Schr\"{o}dinger-like equations for $W^{(n)}_2(y,z)$ and $W^{(n)}_3(y,z)$. Moreover, it is known that $m_2^{(n)}$ and $m_1^{(n)}$ are related to the masses of the vector KK modes, and the wave functions $W^{(n)}_2(y,z), W^{(n)}_3(y,z)$ describe the scalar KK modes. But they appear in \eqref{sch5} and \eqref{sch6} at the same time, which implies that there may be some relationship between the vector KK modes and the scalar ones.

\item
 Equations \eqref{coupl5} and \eqref{coupl6} show a relationship between $W^{(n)}_2(y,z)$ and $W^{(n)}_3(y,z)$.  As Eqs. \eqref{coup3} and \eqref{coup4} read
 \begin{eqnarray}
  m_{\phi}^{(n)2}W_2^{(n)}(y,z)&=&-\partial_y
  \left[
  \partial_y\big(W_2^{(n)}(y,z)\;\text{e}^{-A}\big)
  \text{e}^{2A}\right]\text{e}^{-A},\nonumber\\
  m_{\varphi}^{(n)2}W_3^{(n)}(y,z)&=&-\partial_z
  \left[
  \partial_z\big(W_3^{(n)}(y,z)\;\text{e}^{-A}\big)
  \text{e}^{2A}\right]
  \text{e}^{-A},\nonumber
\end{eqnarray}
we can substitute $W_2^{(n)}(y,z), W_3^{(n)}(y,z)$ into Eqs. \eqref{coupl5} and \eqref{coupl6}, and get
\begin{eqnarray}
  \frac{C_{3}^{(n)}}{2m_{\phi}^{(n)2}}\;\partial_y\big(W_2^{(n)}(y,z)\;\text{e}^{-A}\big)&=&
  \partial_z\big(W_3^{(n)}(y,z)\;\text{e}^{-A}\big),\label{107}\\
  \frac{C_{3}^{(n)}}{2m_{\varphi}^{(n)2}}\;\partial_z\big(W_3^{(n)}(y,z)\;\text{e}^{-A}\big)&=&
  \partial_y\big(W_2^{(n)}(y,z)\;\text{e}^{-A}\big).\label{109}
\end{eqnarray}
It is easily found that
\begin{equation}
  C_{3}^{(n)2}=4m_{\phi}^{(n)2}\;m_{\varphi}^{(n)2}.
\end{equation}
Equations \eqref{107} and \eqref{109} show the relationship between $W^{(n)}_2(y,z)$ and $W^{(n)}_3(y,z)$ more explicitly.

We remember that $W^{(n)}_2(y,z)$ and $W^{(n)}_3(y,z)$ are both related to $W^{(n)}_1(y,z)$ according to \eqref{coupl1}-\eqref{coupl4}. Thus Eqs. \eqref{coupl5} and \eqref{coupl6} also read as
 \begin{eqnarray}
   \frac{C_{3}^{(n)} \text{e}^{2A}}{2C_{1}^{(n)}C_{2}^{(n)}}\;
   \partial_z
  \left[
  \partial_y\big(W_3^{(n)}(y,z)\;\text{e}^{A}\big)
  \text{e}^{-2A}\right]
  &=&\partial_y
  \left[
  \partial_z\big(W_3^{(n)}(y,z)\;\text{e}^{-A}\big)
  \text{e}^{2A}\right] ,\label{w33}\\
 \frac{C_{3}^{(n)} \text{e}^{2A} }{2C_{1}^{(n)}C_{2}^{(n)}}\;\partial_y
  \left[
  \partial_z\big(W_2^{(n)}(y,z)\;\text{e}^{A}\big)
  \text{e}^{-2A}\right]
  &=&
  \partial_z
  \left[
  \partial_y\big(W_2^{(n)}(y,z)\;\text{e}^{-A}\big)
  \text{e}^{2A}\right],\label{w22}
\end{eqnarray}
and Eqs. \eqref{107} and \eqref{109} read as
\begin{eqnarray}
  \frac{C_{3}^{(n)}}{2m_{\phi}^{(n)2}}\;\partial_y
  \bigg(\frac{2}{C_{1}}\partial_z\big(W_1^{(n)}\;\text{e}^{-A}\big)\bigg)&=&
  \partial_z\bigg(\frac{2}{C_{2}}\;\partial_y\big(W_1^{(n)}\;\text{e}^{-A}\big)\bigg),\label{w11a}\\
  \frac{C_{3}^{(n)}}{2m_{\varphi}^{(n)2}}\;\partial_z
  \bigg(\frac{2}{C_{2}}\;\partial_y\big(W_1^{(n)}\;\text{e}^{-A}\bigg)
  &=&
  \partial_y\bigg(\frac{2}{C_{1}}\;\partial_z\big(W_1^{(n)}\;\text{e}^{-A}\big)\bigg).\label{w11b}
\end{eqnarray}
Now, the interesting results can be obtained. From Eqs. \eqref{w33} and \eqref{w22}, we get
\begin{equation}\label{relationc123}
  C_{3}^{(n)}=2\;C_{1}^{(n)}\;C_{2}^{(n)}.
\end{equation}
Then from \eqref{w11a} and \eqref{w11b}, we find
\begin{eqnarray}
m_{\phi}^{(n)2}&=&m_{1}^{(n)2},\label{rela3}\\
m_{\varphi}^{(n)2}&=&m_{2}^{(n)2}.\label{rela4}
\end{eqnarray}

These results imply that the masses of the scalar KK modes are related to those of the vector KK modes. If the vector KK modes are massive, the scalar ones must be also massive. This is different from that in a brane with one extra dimension, where the accompanying scalar KK modes are always massless \cite{Fu2015cfa}.

\end{itemize}
\end{itemize}

\subsection{The orthonormality condition}

We have found the equations for the KK modes, but only the KK modes satisfying the orthonormality conditions \eqref{OrthonormalityCondition} could be localized on the brane. The three orthonormality conditions \eqref{condition1}-\eqref{condition5} are not independent because of the relationships \eqref{coupl2} and \eqref{coupl4}. With the orthonormality condition \eqref{condition1}, the other two orthonormality conditions can be naturally derived. For example,
  \begin{eqnarray}
I^{(nn)}_{3}&\equiv&\frac{1}{2}\int dy dz\; W_2^{(n)}W_2^{(n)}
  =\frac{1}{2}\int dy dz\;\frac{4}{C_{1}^{(n)2}}
  \bigg(\partial_z\big(W_1^{(n)}\;\text{e}^{-A}\big)\bigg)^{2}\;\text{e}^{2A},\nonumber\\
  &&=\frac{1}{2}\int dy dz\;\frac{4}{C_{1}^{(n)2}}W^{(n)}_1(y,z)\;\big[-\partial_z^2+P_2\big]W^{(n)}_1(y,z),\nonumber\\
  &&=\frac{1}{2}\int dy dz\;\frac{4}{C_{1}^{(n)2}}m_{2}^{(n)2}\;W^{(n)2}_1(y,z)=2\;\int dy dz\;W^{(n)2}_1(y,z),
\end{eqnarray}
which is nothing but $I^{(nn)}_{5}$. This means that if there are some localized vector KK modes, there will be some localized scalar ones.

With the relationship between $W^{(n)}_1, W^{(n)}_2$ and $W^{(n)}_3$  and the orthonomality conditions \eqref{OrthonormalityCondition} one can check that if the orthonormality conditions \eqref{OrthonormalityCondition} are satisfied,  the constants $I^{(nn)}_{2}$, $I^{(nn)}_{4}$, $I^{(nn)}_{6}$, $I^{(nn)}_{7}$-$I^{(nn)}_{10}$ are finite, which are consistent with our assumption \eqref{constants}.

\subsection{The gauge invariant effective brane action}

With all above discussions, we now check the gauge invariance of the effective brane action. We have obtained that $I^{(nn)}_{6}=C_1^{(n)}$, $I^{(nn)}_{8}=C_2^{(n)}$, $I^{(nn)}_{10}=2\;C_1^{(n)}\;C_2^{(n)}$, thus the effective action can be rewritten as:
\begin{eqnarray}\label{actioneffective}
  S_{\text{eff}}&=&
     -\frac{1}{4}\sum_n\int d^{4}x \sqrt{-\hat{g}}
     \;\bigg[
     \hat{Y}^{(n)}_{\mu_1\mu_2}\;\hat{Y}_{(n)}^{\mu_1\mu_2}
     +\big(\frac{1}{2}m_1^2+\frac{1}{2}m_2^2\big)
     \;\hat{X}_{\mu_1}^{(n)}\hat{X}^{\mu_1}_{(n)}\nonumber\\
     &&+2\;\partial_{\mu_1}\phi^{(n)}\;\partial^{\mu_1}\phi_{(n)}
     -C_1^{(n)}
     \;\big(\partial_{\mu_1}\phi^{(n)}\;\hat{X}^{\mu_1}_{(n)}
     +\hat{X}_{\mu_1}^{(n)}\;\partial^{\mu_1}\phi^{(n)}\big)\nonumber\\
     &&+2\;\partial_{\mu_1}\varphi^{(n)}\;\partial^{\mu_1}\varphi_{(n)}
     -C_2^{(n)}
     \;\big(\partial_{\mu_1}\varphi^{(n)}\;\hat{X}^{\mu_1}_{(n)}
     +\hat{X}_{\mu_1}^{(n)}\;\partial^{\mu_1}\varphi^{(n)}\big)\nonumber\\
     &&+2\;C_2^{(n)2}\;\phi^{(n)}\phi^{(n)}
     +2\;C_1^{(n)2}\;\varphi^{(n)}\varphi^{(n)}
     -4C_1^{(n)}\;C_2^{(n)}\;\phi^{(n)}\varphi^{(n)}
                                 \bigg],\nonumber\\
    &=&-\frac{1}{4}\sum_n\int d^{4}x \sqrt{-\hat{g}}
     \; \hat{Y}^{(n)}_{\mu_1\mu_2}\;\hat{Y}_{(n)}^{\mu_1\mu_2}\nonumber\\
     &&-\frac{1}{2}\sum_n\int d^{4}x \sqrt{-\hat{g}}
     \;\bigg[
     \big(\partial_{\mu}\phi^{(n)}-\frac{1}{2}\;C_1^{(n)}\;\hat{X}_{\mu}^{(n)}\big)^2
     +\big(\partial_{\mu}\varphi^{(n)}-\frac{1}{2}\;C_2^{(n)}\;\hat{X}_{\mu}^{(n)}\big)^2
     \bigg]\nonumber\\
     &&-\frac{1}{2}\sum_n\int d^{4}x
     \sqrt{-\hat{g}}\;\big(C_2^{(n)}\;\phi^{(n)}-C_1^{(n)}\;\varphi^{(n)}\big)^2.
\nonumber
\end{eqnarray}
We can see that the effective action is invariant under the gauge transformation
\begin{eqnarray}
  \hat{X}_{\mu}&\rightarrow&\hat{X}_{\mu}+\partial_{\mu}\rho,\\
  \phi&\rightarrow&\phi+\frac{1}{2}\;C_1\;\rho, ~~~\varphi\rightarrow\varphi+\frac{1}{2}\;C_2\;\rho,
\end{eqnarray}
where $\rho$ is a scalar field.

\section{Further discussions}

There are some questions needed to be further discussed.
\begin{itemize}
  \item About the ansatz \eqref{line-element1}

  In a model of brane with codimension two, extra dimensions can be assumed to be both compact \cite{Gogberashvili:2007gg}, or one of them is compact and another is non-compact \cite{Collins:2001ni,Cline:2003ak,Costa:2013eua,Liu2007gk,Parameswaran2006db}, or both non-compact \cite{Guo:2008ia}, where the ansatz was supposed as
  \begin{equation}\label{lineelment2}
  ds^2=B(z)^2\big[A(y)^2(\eta_{\mu\nu}dx^\mu dx^\nu+dy^2)+dz^2\big].
  \end{equation}
  The interesting thing is that {in} either the compact case \cite{Gogberashvili:2007gg} or the non-compact case \cite{Guo:2008ia}, the fermion generations can be {obtained}. In this paper, motivated by \eqref{lineelment2} we proposed a similar but more general non-compact metric \eqref{line-element1}. Although in this work, we only study the localization of a bulk $U(1)$ gauge vector field, we plan to investigate the fermion generations like ref.~\cite{Guo:2008ia} in our future work.

  On the other hand, the reason we use this new ansatz \eqref{line-element1} instead of others existing in the literature in codimension 2 is that if we use a metric like \eqref{lineelment2},
  and perform the procedure as before, instead of getting the relationship like \eqref{relationc123}, we will obtain some constraint equations like $\partial_z W_3=0, W_3=0$ and $\partial_z W_2=0, W_2=0$, which will make $W_1=0$. Therefore, in order to avoid the above unreasonable constraint on the wave functions of the KK modes we suppose a more general ansatz \eqref{line-element1}. However, we need further work to investigate how to build the brane with ansatz \eqref{line-element1}. For example, we may consider a complex scalar field and suppose some typical form of \eqref{line-element1} as $A(y,z)=A_1(y)A_2(z)$ or $A(y,z)=A_1(y)+A_2(z)$.

  \item About the applications of our result

  We briefly discuss an application of our result. Begin with the interaction between bulk fermions and gauge boson \cite{Cartas-Fuentevilla:2014sca,Guo:2011qt} in our brane model
   \begin{eqnarray}\label{couplingFB}
     S_I=\int d^4xdydz\sqrt{-g}(-e_6)\bar{\Psi}(x,y,z)\Gamma^M X_M(x,y,z)\Psi(x,y,z),
   \end{eqnarray}
   where $e_6$ is a 6D coupling constant, we can make a dimensional reduction for this action.
   Since there are three types of KK modes (i.e., one type of vector modes and two types of scalar ones) for the bulk $U(1)$ gauge field and one type of fermion KK modes for the bulk fermion, we will have three kinds of interactions between the KK modes in the four-dimensional effective action derived from the fundamental one \eqref{couplingFB}. The coupling between the fermion zero mode and the vector zero mode will recover the usual four-dimensional Coulomb's law, and the couplings between the fermion zero mode and the massive vector KK modes will lead to correction to Coulomb's law \cite{Cartas-Fuentevilla:2014sca,Guo:2011qt}. The Yukawa couplings between the fermion zero mode and the scalar modes will supply the masses of the four-dimensional fermions. The appearance of scalar modes differentiates the present work from the previous, where no scalar modes appear because gauge fixing is usually adopted before KK decomposition.

\end{itemize}

\section{Conclusion}

In this work, we discussed a $U(1)$ gauge vector field on a brane with codimension two. It was found that there are three types of KK modes, one vector and two types of scalars. In many previous papers, the scalar KK modes always were ignored because of some gauge fixing for the bulk vector field. In this paper, we did not choose any gauge for the bulk field, and did a general KK decomposition. We found a gauge invariant effective action on the brane, where the scalar KK modes play an important role.

We first used the general KK decomposition and some orthonormality conditions to obtain the effective action on the brane, which just contains the vector KK modes and two types of scalar ones. Further, by comparing two groups of EOMs, we found the equations of the KK modes, from which the mass spectra and the wave functions can be calculated for a given background solution.

In this paper we only focused on some general discussion of the KK modes, which does not depend on the special solution of the background. Here, we give a simple summary:
\begin{itemize}
  \item The KK modes including the vector and scalar ones satisfy a series of Schr\"{o}dinger-like equations. But these equations are not independent, as there are some relationship between the wave functions of the KK modes;
  \item The masses of the vector KK modes are obtained from the two extra dimensions. The masses of the two types of scalars $m_{\phi}^{(n)}$ and $m_{\varphi}^{(n)}$ are related to the vector ones through $m_{\phi}^{(n)}=m_{1}^{(n)}$ and  $m_{\varphi}^{(n)}=m_{2}^{(n)}$;
  \item The effective brane action is gauge invariant. The vector KK modes couple with the two types of scalar ones, and the two types of scalar KK modes also couple to each other. Because of these couplings the gauge invariance is guaranteed.
\end{itemize}

In fact in the brane world with codimension one, we have found that the effective brane action of the  $1-$form field is gauge invariant \cite{Fu2015cfa}. But the scalar KK modes, which couple with the vector ones, are all massless. Here the brane have one more extra dimension, the scalar KK modes also obtain masses. It is expected that in models with more extra dimensions there will be more types of massive scalar KK modes.


\section{Acknowledgements}

Chun-E Fu sincerely thanks for Bob Holdom for the helpful discussions. This work was supported by the National Natural Science Foundation of China (Grants No. 11405121, No. 11875151, No. 11522541, No. 11305119, No. 11605127, and No. 11374237).


\begin{thebibliography}{10}

\bibitem{ArkaniHamed1998rs}
N.~Arkani-Hamed, S.~Dimopoulos and G.~Dvali, \emph{The hierarchy problem and
  new dimensions at a millimeter}, {\emph{Phys. Lett.} {\bfseries B 429} (1998)
  263} [\href{https://arxiv.org/abs/hep-ph/9803315}{{\ttfamily
  hep-ph/9803315}}].

\bibitem{Randall1999a}
L.~Randall and R.~Sundrum, \emph{A large mass hierarchy from a small extra
  dimension}, \href{https://doi.org/10.1103/PhysRevLett.83.3370}{\emph{Phys.
  Rev. Lett.} {\bfseries 83} (1999) 3370}
  [\href{https://arxiv.org/abs/hep-ph/9905221}{{\ttfamily hep-ph/9905221}}].

\bibitem{Randall1999b}
L.~Randall and R.~Sundrum, \emph{An alternative to compactification},
  \href{https://doi.org/10.1103/PhysRevLett.83.4690}{\emph{Phys. Rev. Lett.}
  {\bfseries 83} (1999) 4690}
  [\href{https://arxiv.org/abs/hep-th/9906064}{{\ttfamily hep-th/9906064}}].

\bibitem{coscon2009}
P.~Dey, B.~Mukhopadhyaya and S.~SenGupta, \emph{Neutrino masses, the
  cosmological constant and a stable universe in a randall-sundrum scenario},
  \href{https://doi.org/10.1103/PhysRevD.80.055029}{\emph{Phys. Rev.}
  {\bfseries D 80} (2009) 055029}
  [\href{https://arxiv.org/abs/0904.1970}{{\ttfamily 0904.1970}}].

\bibitem{ThickBrane2002}
S.~Kobayashi, K.~Koyama and J.~Soda, \emph{Thick brane worlds and their
  stability}, \href{https://doi.org/10.1103/PhysRevD.65.064014}{\emph{Phys.
  Rev.} {\bfseries D 65} (2002) 064014}
  [\href{https://arxiv.org/abs/hep-th/0107025}{{\ttfamily hep-th/0107025}}].

\bibitem{Cline:2003ak}
J.~M. Cline, J.~Descheneau, M.~Giovannini and J.~Vinet, \emph{Cosmology of
  codimension two brane worlds},
  \href{https://doi.org/10.1088/1126-6708/2003/06/048}{\emph{JHEP} {\bfseries
  0306} (2003) 048} [\href{https://arxiv.org/abs/hep-th/0304147}{{\ttfamily
  hep-th/0304147}}].

\bibitem{PhysRevD.66.024024}
A.~Wang, \emph{Thick de sitter 3-branes, dynamic black holes, and localization
  of gravity}, \href{https://doi.org/10.1103/PhysRevD.66.024024}{\emph{Phys.
  Rev.} {\bfseries D 66} (2002) 024024}.

\bibitem{Bazeia2015owa}
D.~Bazeia, A.~S.~L. Jr. and R.~Menezes, \emph{Thick brane models in generalized
  theories of gravity},
  \href{https://doi.org/10.1016/j.physletb.2015.02.037}{\emph{Phys. Lett.}
  {\bfseries B 743} (2015) 98}
  [\href{https://arxiv.org/abs/1502.04757}{{\ttfamily 1502.04757}}].

\bibitem{Archer2011}
P.~R. Archer and S.~J. Huber, \emph{Reducing constraints in a higher
  dimensional extension of the randall and sundrum model},
  \href{https://doi.org/10.1007/JHEP03(2011)018}{\emph{JHEP} {\bfseries 1103}
  (2011) 018} [\href{https://arxiv.org/abs/1010.3588}{{\ttfamily 1010.3588}}].

\bibitem{Gani:2016fqq}
V.~A. Gani, M.~A. Lizunova and R.~V. Radomskiy, \emph{Scalar triplet on a
  domain wall: an exact solution},
  \href{https://doi.org/10.1007/JHEP04(2016)043}{\emph{JHEP} {\bfseries 04}
  (2016) 043} [\href{https://arxiv.org/abs/1601.07954}{{\ttfamily
  1601.07954}}].

\bibitem{Yang:2017evd}
K.~Yang, W.-D. Guo, Z.-C. Lin and Y.-X. Liu, \emph{Domain wall brane in a
  reduced born-infeld-$f(t)$ theory},
  \href{https://doi.org/10.1016/j.physletb.2018.05.017}{\emph{Phys. Lett.}
  {\bfseries B782} (2018) 170}
  [\href{https://arxiv.org/abs/1709.01047}{{\ttfamily 1709.01047}}].

\bibitem{Liu2017gcn}
Y.-X. Liu, \emph{Introduction to extra dimensions and thick braneworlds},
  pp.~211--275.
\newblock 2018.
\newblock \href{https://arxiv.org/abs/1707.08541}{{\ttfamily 1707.08541}}.
\newblock \href{https://doi.org/10.1142/9789813237278_0008}{DOI}.

\bibitem{Zhong2017uhn}
Y.~Zhong, Y.~Zhong, Y.-P. Zhang and Y.-X. Liu, \emph{Thick branes with inner
  structure in mimetic gravity},
  \href{https://doi.org/10.1140/epjc/s10052-018-5527-4}{\emph{Eur. Phys. J.}
  {\bfseries C78} (2018) 45}
  [\href{https://arxiv.org/abs/1711.09413}{{\ttfamily 1711.09413}}].

\bibitem{Parameswaran2009bt}
S.~L. Parameswaran, S.~Randjbar-Daemi and A.~Salvio, \emph{General
  perturbations for braneworld compactifications and the six dimensional case},
  \href{https://doi.org/10.1088/1126-6708/2009/03/136}{\emph{JHEP} {\bfseries
  03} (2009) 136} [\href{https://arxiv.org/abs/0902.0375}{{\ttfamily
  0902.0375}}].

\bibitem{Chang1999nh}
S.~Chang, J.~Hisano, H.~Nakano, N.~Okada and M.~Yamaguchi, \emph{Bulk standard
  model in the randall-sundrum background},
  \href{https://doi.org/10.1103/PhysRevD.62.084025}{\emph{Phys. Rev.}
  {\bfseries D 62} (2000) 084025}
  [\href{https://arxiv.org/abs/hep-ph/9912498}{{\ttfamily hep-ph/9912498}}].

\bibitem{RandjbarDaemi:2000cr}
S.~Randjbar-Daemi and M.~E. Shaposhnikov, \emph{Fermion zero modes on brane
  worlds}, \href{https://doi.org/10.1016/S0370-2693(00)01100-X}{\emph{Phys.
  Lett.} {\bfseries B 492} (2000) 361}
  [\href{https://arxiv.org/abs/hep-th/0008079}{{\ttfamily hep-th/0008079}}].

\bibitem{Oda2001}
I.~Oda, \emph{Localization of bulk fields on ads(4) brane in ads(5)},
  \href{https://doi.org/10.1016/S0370-2693(01)00376-8}{\emph{Phys. Lett.}
  {\bfseries B 508} (2001) 96}
  [\href{https://arxiv.org/abs/hep-th/0012013}{{\ttfamily hep-th/0012013}}].

\bibitem{Mukhopadhyaya:2001fc}
B.~Mukhopadhyaya, S.~Sen and S.~SenGupta, \emph{Bulk torsion fields in theories
  with large extra dimensions},
  \href{https://doi.org/10.1103/PhysRevD.65.124021}{\emph{Phys. Rev.}
  {\bfseries D 65} (2002) 124021}
  [\href{https://arxiv.org/abs/hep-ph/0110308}{{\ttfamily hep-ph/0110308}}].

\bibitem{Ichinose:2002kg}
S.~Ichinose, \emph{Fermions in kaluza-klein and randall-sundrum theories},
  \href{https://doi.org/10.1103/PhysRevD.66.104015}{\emph{Phys. Rev.}
  {\bfseries D 66} (2002) 104015}
  [\href{https://arxiv.org/abs/hep-th/0206187}{{\ttfamily hep-th/0206187}}].

\bibitem{Gogberashvili:2007gg}
M.~Gogberashvili, P.~Midodashvili and D.~Singleton, \emph{Fermion generations
  from 'apple-shaped' extra dimensions},
  \href{https://doi.org/10.1088/1126-6708/2007/08/033}{\emph{JHEP} {\bfseries
  0708} (2007) 033} [\href{https://arxiv.org/abs/0706.0676}{{\ttfamily
  0706.0676}}].

\bibitem{Guo:2008ia}
Z.-q. Guo and B.-Q. Ma, \emph{Fermion families from two layer warped extra
  dimensions}, \href{https://doi.org/10.1088/1126-6708/2008/08/065}{\emph{JHEP}
  {\bfseries 0808} (2008) 065}
  [\href{https://arxiv.org/abs/0808.2136}{{\ttfamily 0808.2136}}].

\bibitem{Liu2008WeylPT}
Y.-X. Liu, L.-D. Zhang, S.-W. Wei and Y.-S. Duan, \emph{Localization and mass
  spectrum of matters on weyl thick branes},
  \href{https://doi.org/10.1088/1126-6708/2008/08/041}{\emph{JHEP} {\bfseries
  0808} (2008) 041} [\href{https://arxiv.org/abs/0803.0098}{{\ttfamily
  0803.0098}}].

\bibitem{Liu:2009mga}
Y.-X. Liu, H.-T. Li, Z.-H. Zhao, J.-X. Li and J.-R. Ren, \emph{Fermion
  resonances on multi-field thick branes},
  \href{https://doi.org/10.1088/1126-6708/2009/10/091}{\emph{JHEP} {\bfseries
  0910} (2009) 091} [\href{https://arxiv.org/abs/0909.2312}{{\ttfamily
  0909.2312}}].

\bibitem{LocalizationWithoutScalar2011JHEP}
A.~Herrera-Aguilar, D.~Malagon-Morejon and R.~R. Mora-Luna, \emph{Localization
  of gravity on a de sitter thick braneworld without scalar fields},
  \href{https://doi.org/10.1007/JHEP11(2010)015}{\emph{JHEP} {\bfseries 1011}
  (2010) 015} [\href{https://arxiv.org/abs/1009.1684}{{\ttfamily 1009.1684}}].

\bibitem{Fu2012sa}
C.-E. Fu, Y.-X. Liu, K.~Yang and S.-W. Wei, \emph{q-form fields on p-branes},
  \href{https://doi.org/10.1007/JHEP10(2012)060}{\emph{JHEP} {\bfseries 10}
  (2012) 060} [\href{https://arxiv.org/abs/1207.3152}{{\ttfamily 1207.3152}}].

\bibitem{Fu2015cfa}
C.-E. Fu, Y.-X. Liu, H.~Guo and S.-L. Zhang, \emph{New localization mechanism
  and hodge duality for $q-$form field},
  \href{https://doi.org/10.1103/PhysRevD.93.064007}{\emph{Phys. Rev.}
  {\bfseries D 93} (2016) 064007}
  [\href{https://arxiv.org/abs/1502.05456}{{\ttfamily 1502.05456}}].

\bibitem{Guo2014nja}
H.~Guo, Q.-Y. Xie and C.-E. Fu, \emph{Localization and quasilocalization of a
  spin-$1/2$ fermion field on a two-field thick braneworld},
  \href{https://doi.org/10.1103/PhysRevD.92.106007}{\emph{Phys. Rev.}
  {\bfseries D 92} (2015) 106007}
  [\href{https://arxiv.org/abs/1408.6155}{{\ttfamily 1408.6155}}].

\bibitem{Vaquera-Araujo2014tia}
C.~A. Vaquera-Araujo and O.~Corradini, \emph{Localization of abelian gauge
  fields on thick branes},
  \href{https://doi.org/10.1140/epjc/s10052-014-3251-2}{\emph{Eur. Phys. J.}
  {\bfseries C 75} (2015) 48}
  [\href{https://arxiv.org/abs/1406.2892}{{\ttfamily 1406.2892}}].

\bibitem{Arun:2016ela}
M.~T. Arun and D.~Choudhury, \emph{Bulk gauge and matter fields in nested
  warping: Ii. symmetry breaking and phenomenological consequences},
  \href{https://doi.org/10.1007/JHEP04(2016)133}{\emph{JHEP} {\bfseries 04}
  (2016) 133} [\href{https://arxiv.org/abs/1601.02321}{{\ttfamily
  1601.02321}}].

\bibitem{Li2017dkw}
Y.-Y. Li, Y.-P. Zhang, W.-D. Guo and Y.-X. Liu, \emph{Fermion localization
  mechanism with derivative geometrical coupling on branes},
  \href{https://doi.org/10.1103/PhysRevD.95.115003}{\emph{Phys. Rev.}
  {\bfseries D 95} (2017) 115003}
  [\href{https://arxiv.org/abs/1701.02429}{{\ttfamily 1701.02429}}].

\bibitem{Mendes2017hmv}
W.~M. Mendes, G.~Alencar and R.~R. Landim, \emph{Spinors fields in co-dimension
  one braneworlds}, \href{https://doi.org/10.1007/JHEP02(2018)018}{\emph{JHEP}
  {\bfseries 02} (2018) 018}
  [\href{https://arxiv.org/abs/1712.02590}{{\ttfamily 1712.02590}}].

\bibitem{Alencar2018cbk}
G.~Alencar, I.~C. Jardim and R.~R. Landim, \emph{$p-$forms non-minimally
  coupled to gravity in randall-sundrum scenarios},
  \href{https://doi.org/10.1140/epjc/s10052-018-5829-6}{\emph{Eur. Phys. J.}
  {\bfseries C78} (2018) 367}
  [\href{https://arxiv.org/abs/1801.06098}{{\ttfamily 1801.06098}}].

\bibitem{Zhou2017bbj}
X.-N. Zhou, Y.-Z. Du, Z.-H. Zhao and Y.-X. Liu, \emph{Localization of
  five-dimensional elko spinors with non-minimal coupling on thick branes},
  \href{https://doi.org/10.1140/epjc/s10052-018-5971-1}{\emph{Eur. Phys. J.}
  {\bfseries C 78} (2018) 493}
  [\href{https://arxiv.org/abs/1710.02842}{{\ttfamily 1710.02842}}].

\bibitem{Zhao2014gka}
Z.-H. Zhao, Y.-X. Liu and Y.~Zhong, \emph{$u(1)$ gauge field localization on
  bloch brane with chumbes-holf da silva-hott mechanism},
  \href{https://doi.org/10.1103/PhysRevD.90.045031}{\emph{Phys. Rev.}
  {\bfseries D 90} (2014) 045031}
  [\href{https://arxiv.org/abs/1402.6480}{{\ttfamily 1402.6480}}].

\bibitem{Zhao2014iqa}
Z.-H. Zhao, Q.-Y. Xie and Y.~Zhong, \emph{New localization method of $u(1)$
  gauge vector field on flat branes in (asymptotic) $ads_{5}$ spacetime},
  \href{https://doi.org/10.1088/0264-9381/32/3/035020}{\emph{Class. Quant.
  Grav.} {\bfseries 32} (2015) 035020}
  [\href{https://arxiv.org/abs/1406.3098}{{\ttfamily 1406.3098}}].

\bibitem{Alencar2014moa}
G.~Alencar, R.~R. Landim, M.~O. Tahim and R.~N. Costa~Filho, \emph{Gauge field
  localization on the brane through geometrical coupling},
  \href{https://doi.org/10.1016/j.physletb.2014.10.040}{\emph{Phys. Lett.}
  {\bfseries B 739} (2014) 125}
  [\href{https://arxiv.org/abs/1409.4396}{{\ttfamily 1409.4396}}].

\bibitem{Zhao2017epp}
Z.-H. Zhao and Q.-Y. Xie, \emph{Localization of $u(1)$ gauge vector field on
  flat branes with five-dimension (asymptotic) ads$_{5}$ spacetime},
  \href{https://doi.org/10.1007/JHEP05(2018)072}{\emph{JHEP} {\bfseries 05}
  (2018) 072} [\href{https://arxiv.org/abs/1712.09843}{{\ttfamily
  1712.09843}}].

\bibitem{Kehagias:2000au}
A.~Kehagias and K.~Tamvakis, \emph{Localized gravitons, gauge bosons and chiral
  fermions in smooth spaces generated by a bounce},
  \href{https://doi.org/10.1016/S0370-2693(01)00274-X}{\emph{Phys. Lett.}
  {\bfseries B 504} (2001) 38}
  [\href{https://arxiv.org/abs/hep-th/0010112}{{\ttfamily hep-th/0010112}}].

\bibitem{Chumbes2011zt}
A.~E.~R. Chumbes, J.~M. Hoff~da Silva and M.~B. Hott, \emph{A model to localize
  gauge and tensor fields on thick branes},
  \href{https://doi.org/10.1103/PhysRevD.85.085003}{\emph{Phys. Rev.}
  {\bfseries D 85} (2012) 085003}
  [\href{https://arxiv.org/abs/1108.3821}{{\ttfamily 1108.3821}}].

\bibitem{Ghoroku:2001zu}
K.~Ghoroku and A.~Nakamura, \emph{Massive vector trapping as a gauge boson on a
  brane}, \href{https://doi.org/10.1103/PhysRevD.65.084017}{\emph{Phys. Rev.}
  {\bfseries D 65} (2002) 084017}
  [\href{https://arxiv.org/abs/hep-th/0106145}{{\ttfamily hep-th/0106145}}].

\bibitem{Freitas2018iil}
L.~F. Freitas, G.~Alencar and R.~R. Landin, \emph{Universal aspects of $u(1)$
  gauge field localization on branes in $d$-dimension},
  \href{https://arxiv.org/abs/1809.07197}{{\ttfamily 1809.07197}}.

\bibitem{Liu2007WeylVo}
Y.-X. Liu, X.-H. Zhang, L.-D. Zhang and Y.-S. Duan, \emph{Localization of
  matters on pure geometrical thick branes},
  \href{https://doi.org/10.1088/1126-6708/2008/02/067}{\emph{JHEP} {\bfseries
  0802} (2008) 067} [\href{https://arxiv.org/abs/0708.0065}{{\ttfamily
  0708.0065}}].

\bibitem{Liu2009uca}
Y.-X. Liu, H.~Guo, C.-E. Fu and J.-R. Ren, \emph{Localization of matters on
  anti-de sitter thick branes},
  \href{https://doi.org/10.1007/JHEP02(2010)080}{\emph{JHEP} {\bfseries 1002}
  (2010) 080} [\href{https://arxiv.org/abs/0907.4424}{{\ttfamily 0907.4424}}].

\bibitem{LocalizationFuPRD2011}
C.-E. Fu, Y.-X. Liu and H.~Guo, \emph{Bulk matter fields on two-field thick
  branes}, \href{https://doi.org/10.1103/PhysRevD.84.044036}{\emph{Phys. Rev.}
  {\bfseries D 84} (2011) 044036}
  [\href{https://arxiv.org/abs/1101.0336}{{\ttfamily 1101.0336}}].

\bibitem{Costa:2013eua}
F.~Costa, J.~Silva and C.~Almeida, \emph{Gauge vector field localization on
  3-brane placed in a warped transverse resolved conifold},
  \href{https://doi.org/10.1103/PhysRevD.87.125010}{\emph{Phys. Rev.}
  {\bfseries D 87} (2013) 125010}
  [\href{https://arxiv.org/abs/1304.7825}{{\ttfamily 1304.7825}}].

\bibitem{Duff2000se}
M.~Duff and J.~T. Liu, \emph{Hodge duality on the brane},
  \href{https://doi.org/10.1016/S0370-2693(01)00520-2}{\emph{Phys. Lett.}
  {\bfseries B 508} (2001) 381}
  [\href{https://arxiv.org/abs/hep-th/0010171}{{\ttfamily hep-th/0010171}}].

\bibitem{Collins:2001ni}
H.~Collins and B.~Holdom, \emph{The randall-sundrum scenario with an extra
  warped dimension},
  \href{https://doi.org/10.1103/PhysRevD.64.064003}{\emph{Phys. Rev.}
  {\bfseries D 64} (2001) 064003}
  [\href{https://arxiv.org/abs/hep-ph/0103103}{{\ttfamily hep-ph/0103103}}].

\bibitem{Liu2007gk}
Y.-X. Liu, L.~Zhao and Y.-S. Duan, \emph{Localization of fermions on a
  string-like defect},
  \href{https://doi.org/10.1088/1126-6708/2007/04/097}{\emph{JHEP} {\bfseries
  04} (2007) 097} [\href{https://arxiv.org/abs/hep-th/0701010}{{\ttfamily
  hep-th/0701010}}].

\bibitem{Parameswaran2006db}
S.~L. Parameswaran, S.~Randjbar-Daemi and A.~Salvio, \emph{Gauge fields,
  fermions and mass gaps in 6d brane worlds},
  \href{https://doi.org/10.1016/j.nuclphysb.2006.12.020}{\emph{Nucl. Phys.}
  {\bfseries B 767} (2007) 54}
  [\href{https://arxiv.org/abs/hep-th/0608074}{{\ttfamily hep-th/0608074}}].

\bibitem{Cartas-Fuentevilla:2014sca}
R.~Cartas-Fuentevilla, A.~Escalante, G.~Germ¨¢n, A.~Herrera-Aguilar and R.~R.
  Mora-Luna, \emph{Coulomb's law corrections and fermion field localization in
  a tachyonic de sitter thick braneworld},
  \href{https://doi.org/10.1088/1475-7516/2016/05/026}{\emph{JCAP} {\bfseries
  1605} (2016) 026} [\href{https://arxiv.org/abs/1412.8710}{{\ttfamily
  1412.8710}}].

\bibitem{Guo:2011qt}
H.~Guo, A.~Herrera-Aguilar, Y.-X. Liu, D.~Malagon-Morejon and R.~R. Mora-Luna,
  \emph{Localization of bulk matter fields, the hierarchy problem and
  corrections to coulomb¡¯s law on a pure de sitter thick braneworld},
  \href{https://doi.org/10.1103/PhysRevD.87.095011}{\emph{Phys. Rev.}
  {\bfseries D 87} (2013) 095011}
  [\href{https://arxiv.org/abs/1103.2430}{{\ttfamily 1103.2430}}].

\end{thebibliography}

\providecommand{\href}[2]{#2}\begingroup\raggedright\endgroup

\end{document}